\documentclass{appolb}
\usepackage{graphicx}

\newcommand{\muu}[1]{\mu_u^{#1}}
\newcommand{\mud}[1]{\mu_d^{#1}}

\begin{document}

\title{SUSY with R-symmetry: confronting EW precision observables and LHC constraints%
\thanks{Presented at the 55th Cracow School of Theoretical Physics, June 20-28, 2015, Zakopane, Poland }%
}
\author{Jan Kalinowski
\address{Faculty of Physics, University of Warsaw, Warsaw, Poland}
}
\maketitle
\begin{abstract}
After motivation and short presentation of the minimal supersymmetric model with R-symmetry
 (MRSSM), we 
address the question of accomodating the measured Higgs boson mass in accordance with 
electroweak precision observables and LHC constraints.  
\end{abstract}
  
\section{Introduction}
This article is based on two recent papers  
by Philip Diessner, Wojciech Kotlarski,  Dominik St\"ockinger and myself~\cite{Diessner:2014ksa,Diessner:2015yna}.

The discovery of a Higgs-boson candidate with a mass near 125 GeV by both ATLAS and CMS Collaborations  at the LHC in 
2012 \cite{Higgsdisc} seemingly completes the Standard Model (SM). 
So far the properties of the new state are within experimental errors 
consistent with the predictions of the SM.  Nevertheless, the true nature of the discovered state has to be  thoroughly 
explored. The run 2 of the LHC (and later the high-lumi run) should clarify if the couplings of the new state are exactly as 
predicted by the SM, whether it unitarizes the WW scattering amplitude, or not, and eventually discover new particles.  
Although the SM is  able to describe a vast number of experimental measurements, 
there are many questions which cannot be address: e.g. 
dark matter, baryogenesis  etc.  In particular, the question of stabilization of the 
Higgs boson mass with respect to the Planck scale has fueled theoretical speculations on beyond the SM physics.  
Among these, the TeV-scale supersymmetry is one of the most theoretically and 
experimentally studied options. So far no direct signal of supersymmetry has been observed by 
the LHC experiments, and only limits on superpartner masses have been derived. However those limits should be taken 
with a grain of salt as  the experimental analyses were performed for simplified models with many additional assumptions.   
The current limits may be not valid in more general supersymmetric scenarios and dedicated phenomenological studies 
are required for non-minimal models. In fact the absence of any direct signal of supersymmetric particle production at the LHC, and the measured Higgs boson mass $\sim$125 GeV close to the upper value of $\sim$135 GeV achievable in the MSSM, are a strong motivation to consider non-minimal SUSY scenarios. $R$-symmetric supersymmetric models, invariant  under a global $U(1)_R$ transformation  $\theta\to e^{i\alpha}\theta$, are particularly well motivated.   $R$-invariance is indeed a symmetry \cite{Fayet} of all basic building blocks of the supersymmetric extension of the  SM.  This symmetry is stronger than R-parity because 
forbids not only baryon- and lepton-number changing terms in the superpotential, as well as  dim-5  operators 
mediating proton decay, but also  removes  left-right sfermion mixing, the   higgsino $\mu$-term and Majorana gaugino 
masses. As a result, several of the most important experimental constraints on supersymmetry are alleviated: contributions 
to CP- and flavor-violating observables can be suppressed even in presence of flavor violation in the sfermion sector, 
and the production cross section for squarks  reduced, making squarks below the TeV scale generically compatible with LHC data.

Since this is the write-up of a lecture given at a school, the next section is devoted to a brief exposition of the MRSSM 
structure.  Then we address the problem of confronting  the MRSSM with the observed Higgs boson mass in accordance 
with electroweak precision measurements.  This is not an obvious task since, as to be seen in the next section,  in the 
MRSSM the lightest Higgs boson tree-level mass is typically reduced compared to the MSSM due to the mixing with 
additional scalars. Moreover, an important MSSM mechanism of generating large radiative corrections due to  the 
stop mixing is absent, and  R-symmetry necessarily introduces an SU(2) scalar triplet, which can increase $m_W$ already 
at the tree level. Nevertheless,  in Refs.~\cite{Diessner:2014ksa,Diessner:2015yna} a number of benchmark points illustrating 
different viable parameter regions have been identified and verified that they are not excluded by further experimental 
constraints from Higgs observables, collider and low-energy physics.  In this write-up we will show the results for only 
one of the benchmarks BMP1 corresponding to $\tan\beta=3$; for other BMP's we refer to our original publications, 
where a comprehensive analysis of the parameter space is discussed and a complete list of references can 
also be found. 
We conclude with summary and outlook.

\section{MRSSM}
Under the global $U(1)_R$ R-symmetry the Grassmann coordinate in the superspace $\{x^\mu,\theta,\bar\theta\}$ transforms as $\theta\to e^{i\alpha}\theta$. By convention we can sign the R-charge 1 ($-1$) to the coordinate $\theta$ ($\bar\theta$). If a  super field $\hat X=\phi_x+\theta\chi+\bar\theta\bar\chi+\ldots$ has a well defined R-charge $R_X$,  so that it transforms as  
$$
\hat X(x^\mu,\theta,\bar\theta) \to e^{i\alpha R_X}\hat X(x^\mu,e^{i\alpha}\theta,e^{-i\alpha}\bar\theta),
$$
the component fields must transform differently. Obviously, the scalar component  $\phi_x$ has R-charge $R_X$, the fermionic component $\chi$ ($\bar\chi$) has R-charge $R_X$$-$1 ($R_X$+1),  etc.  For the gauge invariant kinetic term of a chiral super field $\hat\Phi$ (irrespectively of its R-charge) 
\begin{equation}
{\cal L}\ni \int d^2\theta d^2\bar\theta \; \hat\Phi^\dagger e^{-2g\hat G}\Phi
\end{equation}
to be R-invariant, the gauge vector super field $\hat V$ must be uncharged under R-symmetry.  Since  $\hat V =\ldots -\bar \theta\sigma_{\mu}\theta V^\mu -i \bar\theta\bar\theta\theta\lambda+\ldots$,  the gauge vector field $V^\mu$ is uncharged, while the gaugino $\lambda$ must carry R-charge +1.  Then the kinetic term for the gauge superfields 
\begin{equation}
{\cal L}\ni \int d^2\theta \hat W\hat W,
\end{equation}
where $\hat W$ stands for the the gauge superfield stress tensor $\hat W\ni -i \lambda + \sigma_\mu\bar\sigma_\nu \theta F^{\mu\nu}+\ldots$, 
is also automatically R-invariant. On the other hand,  the Majorana gaugino mass terms $\frac{1}{2}M^M\lambda\lambda $  are forbidden in the soft-SUSY breaking Lagrangian. However, 
Dirac mass terms $M^D\lambda\lambda'$ are perfectly allowed if  additional fermions $\lambda'$ with opposite R-charge  in the adjoint representations of each gauge group factor are introduced.  This can be achieved by introducing gauge chiral-superfield
adjoints $\hat{\cal O},\hat{T},\hat{S}$ corresponding to SU(3)$_c$, SU(2)$_L$ and U(1), respectively, each with R-charge 0. Such a construction amounts to  promoting the  gauge/gaugino sector  to the N=2 supersymmetric structure, which necessarily brings in new scalars, i.e.  for each  group factor, apart from   gauge vector fields and gauge Dirac fermions, there are scalars in the adjoint representations.  The MRSSM therefore contains sgluons -- color-octet scalars,$O$, a scalar SU(2) triplet $T$, and a scalar singlet $S$. 

The asignment of R-charges to the matter chiral superfields is model dependent. In the Minimal $R$-symmetric Supersymmetric Standard Model (MRSSM) \cite{mrssm} it is done in such a way that all SM 
particles have R charge 0 (in analogy to discrete R-parity). Thus the left-chiral  quark and lepton superfields have R-charge 1 and  left-chiral Higgs superfields have R-charge 0.  With this assignment the standard Yukawa terms in  the 
 superpotential are perfectly allowed, while all baryon- and lepton-number violating terms,  as well as dimension-five operators mediating proton decay  are forbidden.  For the same reason the standard Higgs/higgsino $\mu$ term is also forbidden. Therefore, to generate R-symmetric $\mu$ terms (and consequently higgsino mass terms),  the Higgs sector of the MRSSM is extended by adding two iso-doublet superfields $\hat R_u$ and $\hat R_d$  with R-charge 2 (to be called R-Higgs).
 In all, the spectrum of fields in the R-symmetric supersymmetry theory consists of the standard MSSM matter, Higgs and gauge superfields augmented by chiral adjoints $\hat{\cal O},\hat{T},\hat{S}$  and two R-Higgs iso-doublets. The R-charges of the superfields and their component fields are listed in Table~\ref{tab:Rcharges}.  
 
\begin{table}[t]
\begin{center}
\begin{tabular}{c|l|l||l|l|l|l}
\multicolumn{1}{c}{Field} & \multicolumn{2}{c}{Superfield} &
                              \multicolumn{2}{c}{Boson} &
                              \multicolumn{2}{c}{Fermion} \\
\hline 
 \phantom{\rule{0cm}{5mm}}gauge vector    
          & $\hat{g},\hat{W},\hat{B}$        &  $\;\,$ 0 \,
          & $g,W,B$                 & $\;\,$ 0 \,
          & $\tilde{g},\tilde{W}\tilde{B}$             &  +1 \,  \\
matter   & $\hat{l}, \hat{e}^c$                    & $\;$+1 \,
          & $\tilde{l},\tilde{e}^*_R$                 &  +1 \,
          & $l,e^*_R$                                 & $\;\,$\,0 \,    \\
          & $\hat{q},{\hat{d}^c},{\hat{u}^c}$       & $\;$+1 \,
          & $\tilde{q},{\tilde{d}}^*_R,{\tilde{u}}^*_R$ & +1 \,
          & $q,d^*_R,u^*_R$                             & $\;\,$\,0 \,    \\
 $H$-Higgs    & ${\hat{H}}_{d,u}$   & $\;\,$\, 0 \,
          & $H_{d,u}$               & $\;\,$\, 0 \,
          & ${\tilde{H}}_{d,u}$     &  $-$1 \, \\ \hline
\phantom{\rule{0cm}{5mm}} R-Higgs    & ${\hat{R}}_{d,u}$   & +2 \,
          & $R_{d,u}$               & +2 \,
          & ${\tilde{R}}_{d,u}$     & +1 \, \\
  adjoint chiral  & $\hat{\cal O},\hat{T},\hat{S}$     & $\;\,$ 0 \,
          & $O,T,S$                & $\;\,$ 0 \,
          & $\tilde{O},\tilde{T},\tilde{S}$          &  $-$1 \,  \\
\end{tabular}
\end{center}
\caption{The R-charges of the superfields and the corresponding bosonic and
             fermionic components.
        }
\label{tab:Rcharges}
\end{table}

 The MRSSM superpotential takes the following form
\begin{eqnarray}
\nonumber &&W =  \mu_d\,\hat{R}_d \cdot \hat{H}_d\,+\mu_u\,\hat{R}_u\cdot\hat{H}_u\,\,
 - Y_d \,\hat{d}\,\hat{q}\cdot\hat{H}_d\,- Y_e \,\hat{e}\,\hat{l}\cdot\hat{H}_d\, +Y_u\,\hat{u}\,\hat{q}\cdot\hat{H}_u\\ 
 && \phantom{M}+\lambda_d\,\hat{S}\,\hat{R}_d\cdot\hat{H}_d\,+\lambda_u\,\hat{S}\,\hat{R}_u\cdot\hat{H}_u\,
 +\Lambda_d\,\hat{R}_d\cdot \hat{T}\,\hat{H}_d\,+\Lambda_u\,\hat{R}_u\cdot\hat{T}\,\hat{H}_u
\, .
\label{eq:superpot}
 \end{eqnarray} 
Note that the $\Lambda,\lambda$-terms are similar to the usual Yukawa terms, where the $\hat R$-Higgs and $\hat S$ or $\hat T$ play the role of the quark/lepton doublets and singlets. They will turn to be instrumental in achieving the required Higgs boson mass. 

Turning to soft-SUSY breaking, the usual soft mass terms of the MSSM scalar fields are allowed just like in the MSSM.  Similarly,  
the soft  SUSY breaking $B_\mu$, the Higgs, the adjoint scalar  and R-Higgs scalar masses are consistent with R-symmetry. 
 Although  the holomorphic soft mass terms for the adjoint scalars, like $(m^2SS+h.c)$,  
 are also allowed, for simplicity we will neglect them, as well as their 
trilinear couplings among themselves and to the Higgs bosons since their presence does not influence our results significantly. 
On the other hand all trilinear scalar couplings involving Higgs bosons to squarks and sleptons, which in the MSSM are 
notoriously unwanted sources of flavor violation, are removed since they carry non-vanishing R-charge.  Likewise,  
the bilinear coupling of the R-Higgs has  R-charge  4 and therefore is forbidden as well. The $B_\mu$ term is thus 
the only one which destroys the exchange
symmetry between the $H$ and R-Higgs fields. The soft-SUSY breaking scalar mass terms that we take read
\begin{eqnarray}
V_{SB}= &&  B_{\mu}(H_d^- H_u^+- H_d^0 H_u^0 ) + \mbox{h.c.} \nonumber
\\
\, && +m_{H_d}^2 (|H_d^0|^2 + |H_d^-|^2) +m_{H_u}^2 (|H_u^0|^2 + |H_u^+|^2) 
\nonumber  \\ &&
+m_{R_d}^2 (|R_d^0|^2 + |R_d^+|^2)   +m_{R_u}^2 |R_u^0|^2+m_{R_u}^2 |R_d^-|^2 \nonumber
\\ &&
+m_S^2 |S|^2 +m_T^2 |T^0|^2 +m_T^2 |T^-|^2 +m_T^2 |T^+|^2 +m_O^2 |O|^2\nonumber 
\\
  &&+\tilde{d}^*_{L,{i }}  m_{q,{i j}}^{2} \tilde{d}_{L,{j }} 
 +\tilde{d}^*_{R,{i }}  m_{d,{i j}}^{2} \tilde{d}_{R,{j }}  +\tilde{u}^*_{L,{i }} m_{q,{i j}}^{2} \tilde{u}_{L,{j }} +\tilde{u}^*_{R,{i }}  m_{u,{i j}}^{2} \tilde{u}_{R,{j }}
 \nonumber \\ 
 && +\tilde{e}^*_{L,{i}} m_{l,{i j}}^{2} \tilde{e}_{L,{j}} +\tilde{e}^*_{R,{i}} m_{e,{i j}}^{2} \tilde{e}_{R,{j}} +\tilde{\nu}^*_{L,{i}} m_{l,{i j}}^{2} \tilde{\nu}_{L,{j}} \,.
\label{eq:othersoftpot}
\end{eqnarray}
Although the familiar MSSM Weyl fermions $\tilde B, \tilde W,\tilde g$ cannot receive the soft Majorana masses, they can be paired with the corresponding fermionic component of the chiral adjoints $\tilde S, \tilde T, \tilde O$. When the Dirac mass terms are generated by $D$-type spurions, additional terms with auxiliary $\mathcal D$-fields appear in the Lagrangian  \begin{eqnarray}
\nonumber V_D&=&  M_B^D (\tilde{B}\,\tilde{S}-\sqrt{2} \mathcal{D}_B\, S)+
M_W^D(\tilde{W}^a\tilde{T}^a-\sqrt{2}\mathcal{D}_W^a T^a)\\
&&+
M_O^D(\tilde{g}^a\tilde{O}^a-\sqrt{2}\mathcal{D}_g^a O^a)
+ \mbox{h.c.}
\label{eq:potdirac}
\end{eqnarray}
When  the auxiliary fields are eliminated via equations of motion the Dirac mass parameters enter the scalar sector as well. 
 
 Since the R-Higgs fields carry non-vanishing R-charge, they do not develop vacuum expectation values (vev).  The electroweak gauge symmetry breaking is triggered only by  the vev's of neutral EW scalar fields, parameterized as
 \begin{eqnarray} \nonumber 
&H_d^0=  \textstyle{\frac{1}{\sqrt{2}}} (v_d + \phi_d+i  \sigma_{d}), \quad & H_u^0= \textstyle{\frac{1}{\sqrt{2}}} (v_u + \phi_{u}+i  \sigma_{u}) ,\\[-1mm] 
&T^0  =  \textstyle{\frac{1}{\sqrt{2}}} (v_T + \phi_T +i  \sigma_T), \quad    & S   =  \textstyle{\frac{1}{\sqrt{2}}} (v_S + \phi_S +i  \sigma_S)   . 
\end{eqnarray}
The non-vanishing $v_T$  contributes to the $W$ boson mass and shifts the $\rho$ parameter away from one already at tree level. Therefore experimental constraints put an upper limit  $|v_T|\leq 4$ GeV. 

Solving the tadpole equations for $H_d,H_u$, the soft masses  $m^2_{H_d}$ and $m^2_{H_u}$ can be eliminated  
using $v_d$ and $v_u$, and $v^2=v_u^2 + v_d^2$ and $\tan\beta = v_u/v_d$ are defined as in the MSSM.  The other two 
tadpole equations are solved for $v_T$ and $v_S$, allowing us to use $m^2_S$ and $m^2_T$ as input parameters, 
which we assume to be positive to avoid tachyons.  

The neutral scalar fields $(\phi_d,\phi_u,\phi_S,\phi_T)$ mix giving rise to the 4x4 Higgs boson mass matrix. The 2x2 sub-matrix for the $(\phi_d,\phi_u)$ fields takes the same form as in the MSSM, and  for large $m^2_S,m^2_T$ (which we take in the TeV range) the 2x2 sub-matrix for the $(\phi_S,\phi_T$ is approximately diagonal. The 2x2 off-diagonal sub-matrix that mixes the two sectors  reads as
\begin{eqnarray}
\mathcal{M}_{21}= \left(
\begin{array}{cc}
 v_d ( \sqrt{2}\lambda_d \mud{+} -g_1 M_B^D )\; & \;
v_u (\sqrt{2} \lambda_u\muu{-} +g_1 M_B^D) \\[2mm]
v_d ( \Lambda_d \mud{+} + g_2 M_W^D) \;& - 
 v_u (\Lambda_u  \muu{-} + g_2 M_W^D) \\
\end{array} \right),
\end{eqnarray}
where the effective $\mu$-parameter is given by $\mu_i^{\pm}
=\mu_i+\frac{\lambda_iv_S}{\sqrt2}
\pm\frac{\Lambda_iv_T}{2}$, $i=u,d$.

In general, the mixing between  $\phi_d,\phi_u$ and $\phi_S,\phi_T$ leads to a reduction of the lightest tree-level Higgs boson mass compared to the MSSM.   In \cite{Diessner:2014ksa} an approximate formula has been derived 
\begin{equation}
m_{H_1}^2 \approx \left(m_Z^2  - v^2 (
\frac{(g_1 M^D_B+\sqrt{2}\lambda\mu)^2}{4(M^D_B)^2+ m_S^2}
+
\frac{(g_2 M^D_W+\Lambda\mu)^2}{4(M_W^D)^2+ m_T^2})
\right) \cos^2 2\beta\;,
\label{eq:approx_treehiggs}
\end{equation}
in the limiting case of large $m_A^2$ and 
assuming $\lambda=\lambda_u=-\lambda_d$, $\Lambda=\Lambda_u=\Lambda_d$, $\mu_u=\mu_d=\mu$ and $v_S= v_T=0$.
It is clear that the MSSM upper limit of $m_Z$ at tree level can be substantially reduced by terms depending on the new model parameters. Therefore in the MRSSM loop corrections must play even more significant role, which we discuss in the next section. 

\section{Loop-corrected Higgs boson masses}
The parameters of the model are renormalized 
in the  $\overline{\rm{DR}}$ scheme
and $v_d$, $v_u$, $v_S$ and $v_T$ are given by the minimum of the 
loop-corrected effective potential. 
The pole mass $m^2_{\rm{pole}}$ of a field  is given by the pole
of the full propagator 
\begin{equation}
\label{eq:PropPole}
0 {=} \det\left[p^2 \delta_{ij}- \hat{m}_{ij}^2 + \Re(\hat{\Sigma}_{ij}(p^2))\right]_{p^2=m^2_{\rm{pole}}}\;,
\end{equation}
where $p$ is the momentum, $\hat{m}^2$ the tree-level mass matrix and $\hat{\Sigma}(p^2)$ 
the finite part of the self-energy corrections. The one-loop self energies have been computed exactly using 
\texttt{FeynArts}~\cite{FA}, \texttt{FormCalc} \cite{FC} and Feynman rules generated by \texttt{SARAH} \cite{SA} 
properly modified to match our model.  
Since the above equation cannot be solved analytically for $m^2_{\rm pole}$, the solution has to be found numerically. 
With \texttt{SARAH} an MRSSM version of the \texttt{SPheno} spectrum generator \cite {SP} has been created to 
calculate the mass spectrum at full one-loop level.  The results have been checked with a recent framework \texttt{FlexibleSUSY}
\cite{FlexibleSUSY}.
 
Before presenting numerical results of full one-loop calculations, it is instructive to  discuss the self energies in the 
effective potential approach.  In the MSSM the dominant one-loop contribution to the Higgs mass matrix comes from the top/stop sector. In the MRSSM they are also important but because of the absence of stop mixing they are simpler 
$$\Delta m^2_{H_1}\sim \frac{6v^2}{16\pi^2} Y_t^4\log\frac{m_{\tilde t_1}m_{\tilde t_2}}{m^2_t}$$
and, as a result, for the same stop mass cannot reach the value as high as in the MSSM. Since the MRSSM superpotential contains new $\lambda_{u,d}$ and $\Lambda_{u,d}$ terms with  a Yukawa-like structure, one can expect additional corrections proportional 
to $\lambda^4,\Lambda^4$ and logarithms of soft masses.    
Using the same approximation as in eq.\ref{eq:approx_treehiggs}, the lightest Higgs 
state is given mainly by $\phi_u$ and only $(\phi_u,\phi_u)$ component of the mass matrix needs to be computed and simple analytical expressions can be derived.  For example, the $\lambda^4$ term gives the following contribution
\begin{equation}
\label{eq:1loop}
\Delta m^2_{H_1}\sim \frac{2v^2}{16\pi^2} \lambda^4\log\frac{M_{R_u}m_S}{(M_B^D)2}
\end{equation}
with a similar structure of the $\Lambda^4$ and somewhat more complicated for the $\lambda^2\Lambda^2$ terms.

\begin{figure}[th]
\centering
\includegraphics[width=6cm]{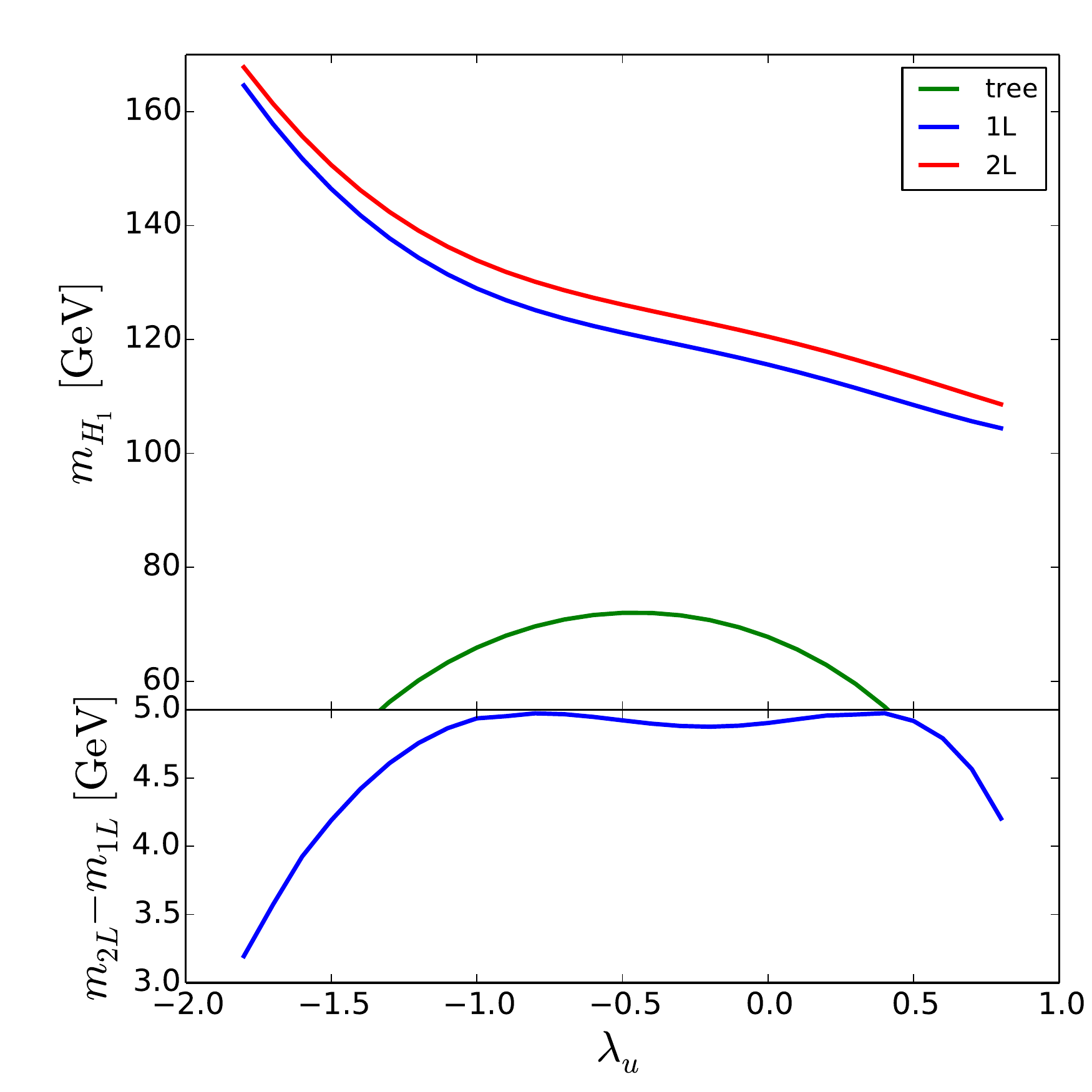}\includegraphics[width=6cm]{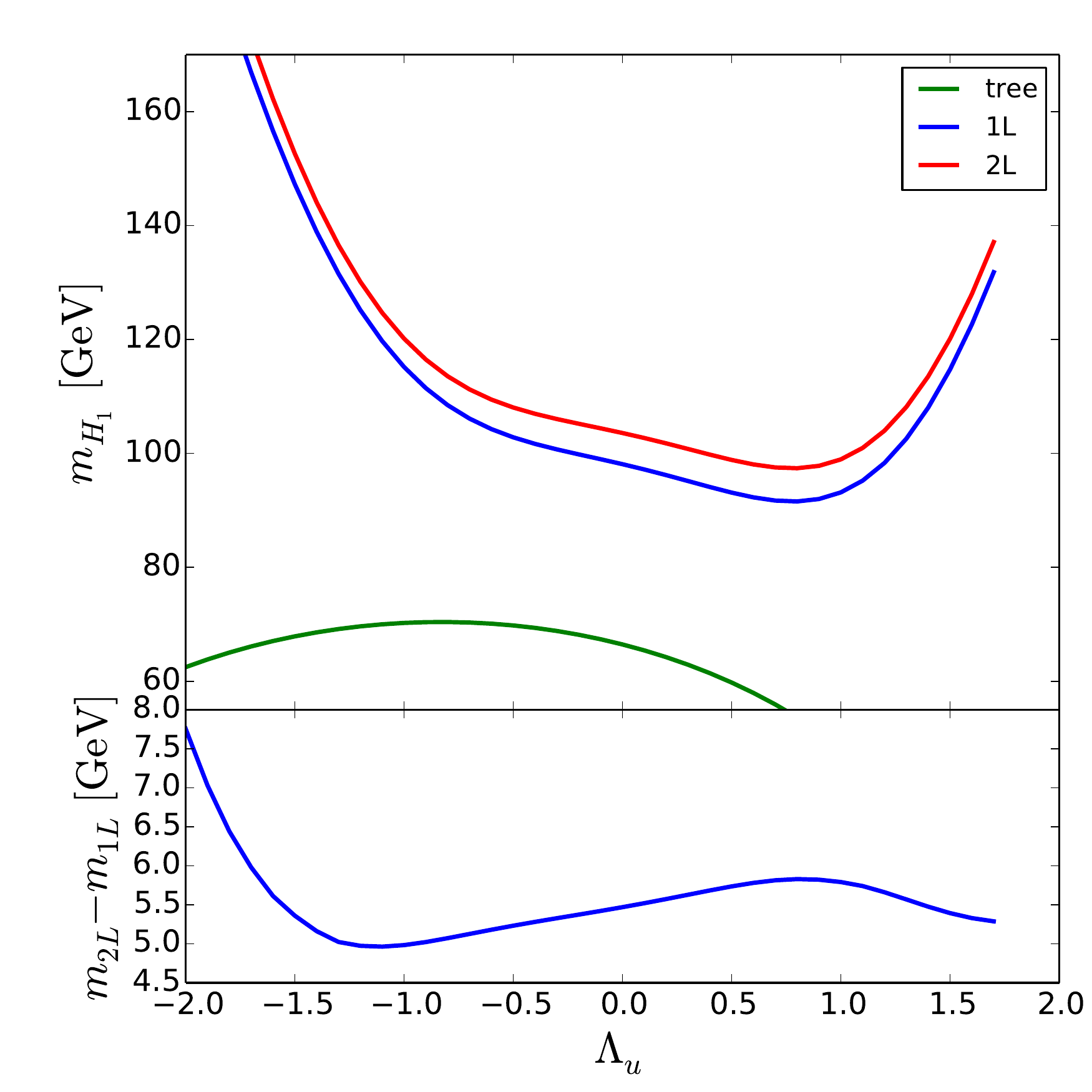}
\caption{Lightest MRSSM Higgs boson mass $m_{H_1}$, and the difference $m_{2L}-m_{1L}$ between masses calculated at the two-loop and one-loop level, as a function of $\lambda_u$, 
$\Lambda_u$, respectively.  In the upper parts of the figure lines from top to bottom correspond to two-loop, one-loop and tree level calculations. Other parameters are set to the values of benchmark point BMP1 with $\tan\beta=3$. [from Ref.\cite{Diessner:2015yna}]}
\label{img:comp}
\end{figure}

For numerical analyses of one-loop corrections no approximation is used and full dependencies on the parameters 
are taken into account.  In \cite{Diessner:2014ksa} three representative benchmarks BMP1, BMP2, BMP3 have been 
identified with $\tan\beta=3,\, 10,\, 40$, respectively, for which the Higgs boson mass can be met in accordance 
with LHC constraints, and with 
EW precision observables,  to be discussed in the  next section.  The dependence of the 
lightest Higgs boson mass calculated at  tree, one-loop (and two-loop) levels for one of the benchmarks, 
the BMP1, is shown in Fig.\ref{img:comp} as a function of one of the parameters, 
with all others set to the benchmark 
values.  As already mentioned, the tree-level mass is significantly reduced below $m_Z\cos\beta$ for large values 
of $\lambda_u,\Lambda_u$, as the mixing between doublets and the singlet and the triplet gets exchanced. The dependence 
on $\lambda_d,\Lambda_d$ is significantly weaker since the lightest Higgs gets dominant contribution from $\phi_u$ even 
for a low $\tan\beta=3$.  The size of one-loop top/stop Yukawa contribution alone can be judged from 
the value read at $\Lambda_u=\lambda_u=0$, since 
then only stop/top contributions are significant. Evidently, 
a stop mass of 1 TeV, as set in the benchmark points, is not 
enough in the MRSSM to achieve the correct value of the Higgs boson mass, due to the absence of left-right mixing.
The full one-loop result shows large positive contributions 
from $\lambda,\Lambda$ terms.  Although the tree level result falls quadratically with $\lambda,\Lambda$, as expected from Eq.~\ref{eq:approx_treehiggs}, the  one-loop result shows quartic dependence, as seen e.g.\ from 
Eq.\ref{eq:1loop}, which explains the behavior of the sum.   Thus the $\lambda,\Lambda$ one-loop 
contributions can push the Higgs boson mass  to the measured value for values  of $\lambda_u,\Lambda_{u}$ close to unity.

Since the one-loop corrections are large, the question arises about the size of higher-order corrections. 
In Ref.~\cite{Diessner:2014ksa} an estimate of higher-order corrections has been given with a conclusion that an 
expected two-loop contribution for the lightest Higgs boson mass should not exceed 6 GeV. This estimate has 
been verified in Ref.~\cite{Diessner:2015yna} using the recently updated \texttt{SARAH} code \cite{newSARAH} that provides \texttt{SPheno} 
routines to calculate two-loop corrections  in the effective potential approach and the gauge-less limit $g_{1,2}=0$. 

At two-loops, the $\lambda,\Lambda$ corrections  should behave in a manner similar to the pure top/stop
 two-loop contributions in the MSSM without stop mixing. And in fact, their numerical impact turns to be rather 
 small, typically below 1 GeV, unless the couplings $\lambda,\Lambda$ become very large, $|\lambda, \Lambda|\gg1$ . 
 But at two-loops  also 
  the strongly interacting sector  and strong coupling $\alpha_s$ enter directly to the Higgs boson mass 
predictions and these corrections can be expected to be sizable.  Apart from the gluon, 
they involve the Dirac gluino and the sgluon, the scalar component  of 
the octet superfield $\hat{\cal O}$, which depend on the sgluon soft mass parameter $m_O$  
and the Dirac mass $M_O^D$. Note that $M_O^D$ appears not only directly as the gluino mass but, via Eq.~(\ref{eq:potdirac}), also in couplings and mass terms of sgluons. In particular, it causes the splitting between the real and imaginary parts of the complex sgluon field $O= {\textstyle\frac{1}{\sqrt{2}}}(O_S+i O_A)$, with tree-level masses $m^2_{O_S} = 4 (M_O^D)^2 + m^2_{O}$ and $m_{O_A}^2=m^2_{O}$.  

 Compared to the MSSM, there are 
important differences due to the Dirac nature of the gluino, the lack of left-right stop mixing and the vanishing 
$\mu$-parameter. For example, the diagrams with 
fermion mass insertions, corresponding to $\overline{FF}S$-type contributions in the notation of 
Ref.~ \cite{Martin:2001vx}, are not present 
in the MRSSM due to the absence of L-R mixing between squarks (for a comprehensive discussion of similarities  
and differences of two-loop results in the MSSM and MRSSM, 
see Ref.~\cite{Diessner:2014ksa}). 
 
\begin{figure}[th]
\centering
\includegraphics[width=6cm]{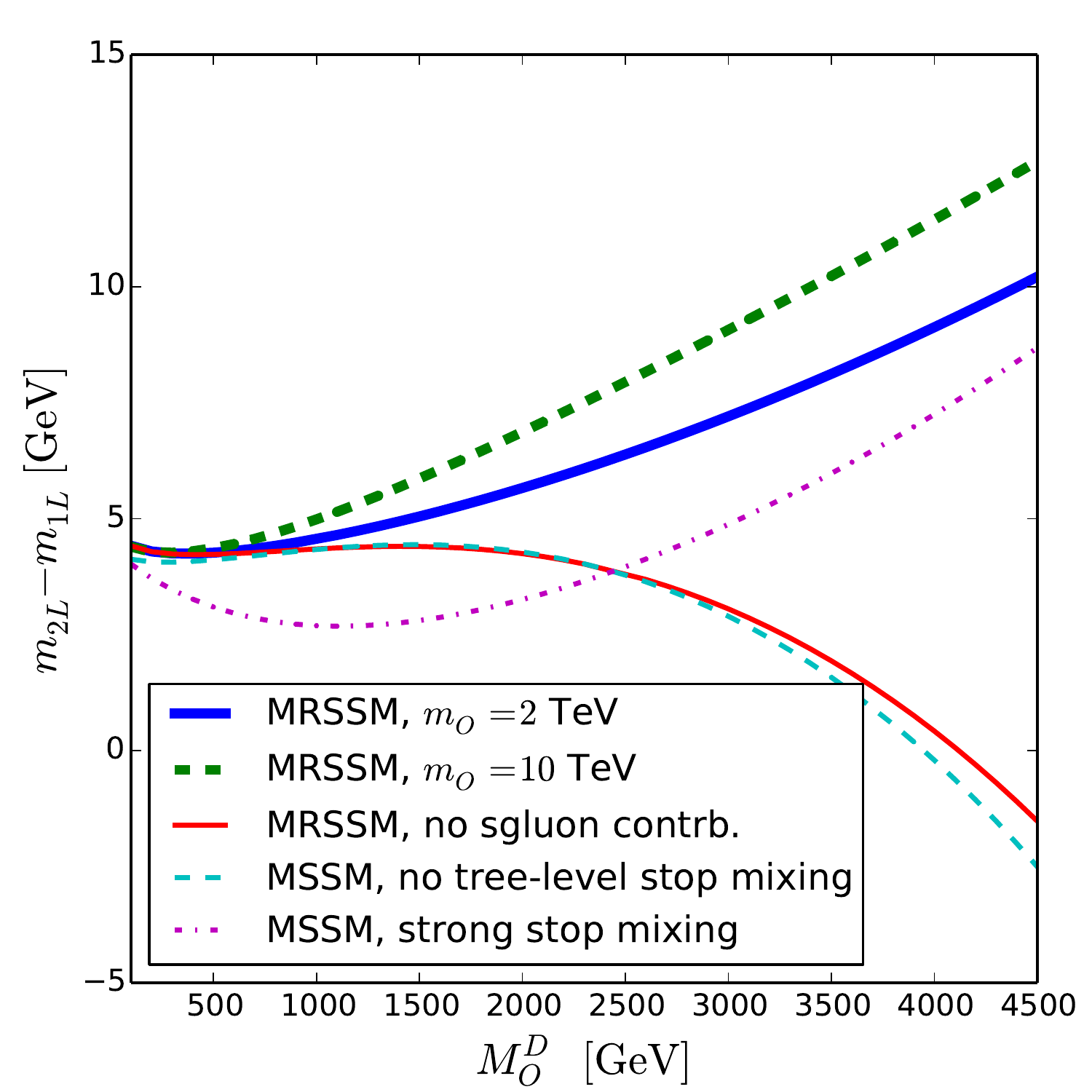}
\caption{Gluino mass dependence of two-loop contributions to the lightest Higgs boson mass  in the 
MRSSM for BMP1. The results are shown for  two different values of the soft sgluon mass parameter 
$m_O=2$~TeV (thick solid blue line) and $10$~TeV (thick dashed green line) with all contributions, respectively, and without
the sgluon contributions (thin solid red line). 
For comparison also the MSSM contributions for no (thin dashed light
blue line) and maximal (purple dotted line) stop mixing are plotted. [from Ref.\cite{Diessner:2015yna}]}
\label{img:2loop}
\end{figure}
 
Fig.~\ref{img:2loop} shows the gluino mass dependence of the complete  two-loop correction to the lightest Higgs boson mass. Curves are drawn for two different values of the sgluon mass parameter $m_O=2$ and 10 GeV  with all other parameters 
corresponding to BMP1. For comparison, the two-loop result without the sgluon contribution is shown as well, and the MSSM result 
with strong stop mixing and without any sfermion mixing the at tree-level. 

In the MSSM without sfermion mixing the
gluino contribution is precisely the same as in the MRSSM since 
the Dirac or Majorana nature of the gluino does not matter as the Dirac partner, 
the octet superfield $\hat{O}$, has no direct couplings to quark superfields. 
This explains why the lines for the MRSSM without sgluon and the MSSM without stop mixing have 
very similar $M^D_O$ dependence.  

Including the sgluon diagram in the MRSSM 
strongly changes the behavior.  Surprisingly, the full MRSSM two-loop
contributions resemble the MSSM contributions with large stop
mixing (corresponding to $X_t=2000$ GeV), however for different reasons. In the MSSM the increase is  due to the
additional  $\overline{FF}S$-type diagram which is directly
proportional to gluino mass, while in the MRSSM,  the sgluon diagram
grows with $M_O^D$, both due to the sgluon-stop-stop coupling, which
scales like $M_O^D$, 
 and to an increase of the  sgluon mass $m_{O_S}$.
With the sgluon contributions the total two-loop contributions
to the Higgs boson mass in the MRSSM are larger than the ones in the
MSSM. They are further increased by heavy sgluons. With the positive two-loop correction a somewhat 
smaller value of the $\Lambda_u$ is needed to meet the experimentally measured Higgs boson mass. 
 
 Overall, the two-loop contribution amounts to approximately +5 GeV, in agreement with previous estimate, and confirms the validity of the perturbative expansion in spite of the large one-loop result. 
 
\section{Electroweak observables} 
 
Since the non-vanishing $v_T$ of the scalar triplet contributes to the tree-level $W$ mass and shifts the $\rho$ parameter 
from 1,  it is constrained to be small: for all our benchmarks it is below 1 GeV. Small $v_T$ implies, 
through tadpole equations, a large 
value of the triplet soft mass and consequently somewhat split spectrum of Higgs bosons. In the SM and the MSSM 
the top Yukawa coupling dominates loop corrections to $m_W$. Therefore it should be expected that due to their  Yukawa-like character, the $\lambda,\Lambda$ couplings will also contribute at loop-level to electroweak observables (EWO), in particular to the $W$ boson mass. 

Beyond tree-level the $W$ boson mass   can be obtained from the precisely measured muon decay constant using 
(hats denote $\overline{DR}$-renormalized quantities in the MRSSM)
\begin{equation}
m_W^2 = \frac{1}{2} m_Z^2 \hat{\rho} \left [ 1 + \sqrt{1
- \frac{4 \pi \hat{\alpha}}{\sqrt{2} G_\mu m_Z^2 \hat{\rho}
(1-\Delta \hat{r}_W)}}\;\right ] \label{w-mass-master-formula2},
\end{equation}
where $\hat \rho$ contains only oblique and $\Delta \hat{r}_W$
both oblique and non-oblique corrections which depend on the entire
particle content of the model.  The above formula also
properly resums leading two-loop SM corrections \cite{degrassi}. 
The  numerical calculation of $\hat{\rho}$ and $\Delta \hat{r}_W$ has been performed with the help of \texttt{SARAH} 
appropriately modified to account for the triplet scalar contribution.

It is convenient to rewrite the one-loop approximation to the $W$ boson mass it terms of the electroweak 
precision parameters $S$, $T$ and $U$ as 
\begin{equation}
m_W= m_W^{\rm{ref}} +\frac{\hat{\alpha} m_Z \hat{c}_W}{2(\hat{c}_W^2-\hat{s}_W^2)}\left (-\frac{S}{2}+\hat{c}^2_WT+\frac{\hat{c}^2_W-\hat{s}^2_W}{4 \hat{s}_W^2}U \right)\;,
\label{eq:mW-STU}
\end{equation}
where $\hat{c}^2_W=1-\hat{s}^2=m^2_W/m^2_Z\hat\rho$, and  $m_W^{\rm ref} $ is the $W$ mass calculated in the SM. The advantage of computing $S,T,U$ parameters is that they can be used in the calculation of several EWO.  
The main contribution to $m_W$ from the MRSSM sector can be described in terms of  the $T$ parameter. It receives input from three sectors: charginos/neutralinos, Higgs and R-Higgs bosons. The contribution from R-Higgses has a similar structure to the stop/sbottom. Since in our benchmarks soft masses $m^2_{R_u}$ and $m^2_{R_d}$ are large, the mixing between R-Higgses and mass-splitting is small leading to negligible contribution to $T$. The contribution from the triplet scalar is suppressed by the large soft triplet mass $m^2_T$. The dominant contribution thus comes from the chargino/neutralino sector. For example, in the simplifying case of $\lambda_u=g_1$ and $\mu_u=M_W^D$, it  can be written as
\begin{equation}
T=\frac{1}{16{\hat s}^2_W{\hat m}^2_W}\frac{v_u^4}{(M^D_W)^2}\times ({\rm 4^{th}~order~polynomial~in~} g_2,\Lambda_u).
\end{equation}
  
\begin{figure}[th]
\centering
\includegraphics[width=0.45\textwidth]{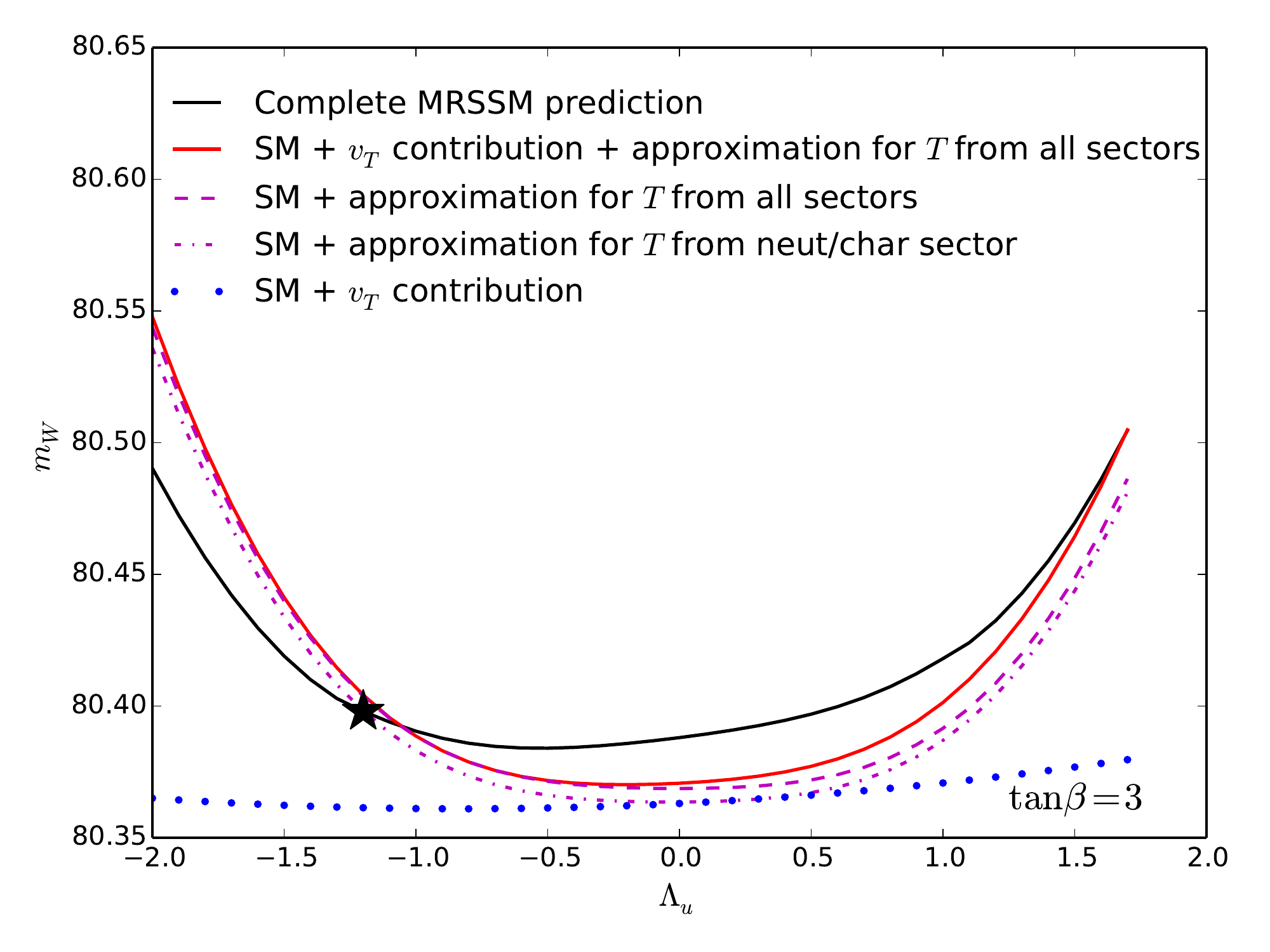}
\caption{The  W boson mass as a function of $\Lambda_u$, calculated using full MRSSM contributions and different approximations
for the $T$-parameter for BMP1 (marked by the black star.}
\label{img:mwlam}
\end{figure}

Figure \ref{img:mwlam}
shows the $\Lambda_u$ dependence of the full calculation of the $W$ mass in the MRSSM (black solid line), as well as 
in various approximations. Other lines   in the figure contain the
full SM contribution, but the MRSSM contributions are
taken into account either completely, or only via the $T$-parameter in
various approximations, or from the tree-level triplet vev
contribution. The figure shows that the chargino/neutralino approximation already gives an excellent approximation to
the full $T$-parameter. The $T$-parameter, together with the
tree-level triplet vev contribution, provides a good approximation to
the full result. The remaining difference from non-$T$-parameter
oblique corrections, vertex and box contributions, and leading higher
loop contributions, is within $\pm20$~MeV, except for
$|\Lambda_u|\gg1.5$.

\section{Conclusions and outlook}
In my talk I have discussed the structure of the R-symmetric supersymmetric  extension of the Standard Model 
and recent progress in the precision calculation of the Higgs and $W$ boson masses. Compared to the MSSM, 
the model contains new states: R-Higgs SU(2)-doublets and singlet, SU(2)-triplet and SU(3)-octet superfields, 
whose fermionic components allow us to write down the Dirac mass terms for gauginos and higgsinos.

We have seen that one can acomodate the observed Higgs boson mass in accordance with precision observables. 
The experimental values of $m_{H_1}$ and $W$ impose stringent and non-trivial constraints on the parameter 
space of the model. Nevertheless, it is easy to identify regions in the parameter space which acomodate the 
measured values and are in accordance with experimental data, as checked explicitly with \texttt{HiggsBounds}  \cite{HB4} 
and \texttt{HiggsSignals} \cite{HS}, as well as selected low-energy flavor constraints.  We have computed the full one-loop 
corrections to  both $m_{H_1}$ and $W$, and the two-loop correction to the Higgs mass in the effective potential 
approach. Numerical calculations have been cross-checked with analytic calculations of the most important new 
corrections. We have found that  large scalar masses are favorable, of order 1$-$3 TeV. The resulting large mass 
ratios enhance loop-corrections to the lightest Higgs boson mass and suppress contribution from new states 
to the $W$ boson mass.  Most instrumental are the new superpotential couplings $\lambda,\Lambda$, which 
play a role similar to the top/bottom Yukawa couplings with R-Higgses and singlet/triplet replacing quark doublets 
and singlets.  With $\lambda,\Lambda$ of order 1, like the top Yukawa coupling,  the Higgs boson mass 
of $\sim$125 GeV can easily be obtained even for top squarks below 1 TeV in spite of lack of L-R sfermion mixing,

The proposed benchmark points have many of the new states within the reach of the run-2 of the LHC, in 
particular the supersymmetric fermions. It would be extremely exciting to see some of them in the current 
run of the LHC.

So far we have exploited scenarios in which the lightest Higgs boson  is the SM-like. In such cases the mixing 
with new states lowers the tree-level $m_{H_1}$ compared to the MSSM value, calling for even larger 
loop corrections to meet the   measured value. However, one can contemplate an alternative scenario in 
which the lightest Higgs boson is mostly singlet, and the next one is the SM-like. 
In such a case the second-lightest Higgs state gets pushed up via mixing already at tree-level, thereby reducing the 
required loop corrections \cite{nextone}.  Similar scenarios have been 
considered in the next-to-minimal MSSM.  

\section*{Acknowledgments}
I would like to thank Philip Diessner, Wojciech Kotlarski and Dominik St\"ockinger for a very fruitful collaboration. 
Work supported in part by the Polish National Science Centre
grants under OPUS-2012/05/B/ST2/03306, DEC-2012/05/B/ST2/02597,
and  the European Commission through the contract PITN-GA-2012-316704 (HIGGSTOOLS)


\end{document}